\documentclass[13 pt]{article}
\usepackage{amssymb}

\newcommand{\beq}{\begin{equation}}
\newcommand{\eeq}{\end{equation}}
\newcommand{\bea}{\begin{eqnarray*}}
\newcommand{\eea}{\end{eqnarray*}}
\newcommand{\beqa}{\begin{eqnarray}}
\newcommand{\eeqa}{\end{eqnarray}}

\begin{document}

\newfont{\elevenmib}{cmmib10 scaled\magstep1}%

\newcommand{\Title}[1]{{\baselineskip=26pt \begin{center}
            \Large   \bf #1 \\ \ \\ \end{center}}}
\hspace*{2.13cm}%
\hspace*{1cm}%
\newcommand{\Author}{\begin{center}\large
           Pascal Baseilhac\footnote{
baseilha@phys.univ-tours.fr} and Kozo Koizumi\footnote{kozo.koizumi@lmpt.univ-tours.fr}
\end{center}}
\newcommand{\Address}{{\baselineskip=18pt \begin{center}
           \it Laboratoire de Math\'ematiques et Physique Th\'eorique CNRS/UMR 6083,\\
Universit\'e de Tours, Parc de Grandmont, 37200 Tours, France
      \end{center}}}
\baselineskip=13pt

\bigskip
\vspace{-1cm}

\Title{A new (in)finite dimensional algebra for quantum integrable models}\Author

\vspace{- 0.1mm}
 \Address

\vskip 0.6cm

\centerline{\bf Abstract}\vspace{0.3mm}  \vspace{1mm}
A new (in)finite dimensional algebra which is a fundamental dynamical symmetry of a large class 
of (continuum or lattice) quantum integrable models is introduced and studied in details. Finite dimensional 
representations are constructed and mutually commuting quantities - which ensure the integrability of the system
 - are written in terms of the fundamental generators of the new algebra. Relation with the deformed Dolan-Grady 
integrable structure recently discovered by one of the authors and Terwilliger's tridiagonal algebras is described. 
Remarkably, this (in)finite dimensional algebra is a ``$q-$deformed'' analogue of the original Onsager's 
algebra arising in the planar Ising model.  Consequently, it provides a new and alternative algebraic framework for studying massive, as well as conformal, quantum integrable models. 

\vspace{0.1cm}  {\small PACS: 02.20.Uw; 11.30.-j; 11.25.Hf;
11.10.Kk}
\vskip 0.8cm

\vskip -0.6cm

{{\small  {\it \bf Keywords}: Onsager's algebra; Tridiagonal algebra; Dolan-Grady relations; Quadratic
algebras; Integrable models}}
%
%

\section{Introduction}
Two-dimensional completely integrable models (continuum or lattice) are characterized by the existence of an (in)finite set of mutually commuting conserved quantities. Such property allows to solve the model exactly without considering approximation schemes. Since the exact solution of the planar Ising model by Onsager in 1944 \cite{Ons44}, several methods have been proposed to analyse in a nonperturbative way integrable models. For instance, the factorized scattering theory (based on Yang-Baxter/star-triangle relations), quantum group symmetry, Bethe ansatz techniques and conformal field theory framework are constantly applied to derive exact results (such as the exact $S-$matrix, VEVs, forms factors,...) in quantum integrable models. For systems with an (in)finite number of degrees of freedom, integrability takes its roots in the existence of an (in)finite dimensional symmetry.  
For instance, in the context of (massless) conformal field theory, the infinite dimensional Virasoro algebra with fundamental generators $L_n$ satisfying
\beqa
[L_n,L_m] = (n-m)L_{n+m} + \frac{c}{12}(n^3-n)\delta_{n+m,0}\ ,\label{Vir}
\eeqa
where $c$ denotes the central charge, actually gives a powerful algebraic approach to critical statistical systems and corresponding field theories \cite{BPZ84}. Indeed, exact results like correlation functions (which were previously not accessible using standard techniques like renormalization group methods) can be derived in a systematic manner using the properties of the Virasoro algebra. Later on, other types of infinite dimensional conformal symmetries including for instance supersymmetry \cite{Frie85}, parafermionic symmetry \cite{Paraf}, ${\cal W}-$algebras \cite{W} and current algebras \cite{current} have been considered, extending (\ref{Vir}).
Similarly, they found several applications in a large class of critical statistical systems or conformal field theories with enlarged symmetries.

In the context of integrable massive quantum field theory or lattice systems, hidden symmetries associated with quantum groups \cite{Ber91,Miwa} or deformed Virasoro algebra \cite{Lukya96} (in the vicinity of critical points) have been introduced in order to derive exact results in these models. However, an infinite dimensional algebra characterizing the dynamical symmetry of a large class of quantum integrable {\it massive} models has not yet been found. The only known example in this direction is the Onsager's algebra with generators $A_k,G_l$ satisfying
\beqa
[A_k,A_l]=4G_{k-l}\ , \quad [G_l,A_k]= 2A_{l+k}-2A_{-l+k}\ , \quad [G_k,G_l]=0\ \label{Ons}
\eeqa
for any integers $k,l$. This infinite dimensional Lie algebra was originally introduced in \cite{Ons44} in order
 to solve the planar Ising model in zero magnetic field. Although it played a crucial role in the original solution of the Ising model, this algebra only arises in a few other quantum integrable models (XY, superintegrable chiral Potts \cite{Potts} and generalizations \cite{Ahn}). In these models, all conserved quantities ${I}_{2k+1}$ can be simply expressed in 
terms of the fundamental generators $A_k$ as
\beqa
{I}_{2k+1}= \frac{\kappa}{2}({A_k+A_{-k}}) +  \frac{\kappa^*}{2}({A_{k+1}+A_{-k+1}})\ , \label{comOns}
\eeqa
for $k\geq 0$ where $\kappa,\kappa^*$ are arbitrary parameters. Also, note that in the 80s the Onsager's algebra was shown to be closely related with the integrable structure discovered by Dolan and Grady in \cite{DG}. Despite of its nice properties, the Onsager's algebra remained an interesting curiosity in the last sixty years.\vspace{1mm} 

Clearly, identifying the underlying (in)finite dimensional symmetry in quantum integrable systems is a fundamental and important problem that we wish to adress in this paper. Indeed, we construct explicitly and study in details an (in)finite dimensional algebraic structure with fundamental generators ${\textsf W}_{-k}$, ${\textsf W}_{k+1}$, ${\textsf G}_{k+1}$, ${\tilde{\textsf G}}_{k+1}$ satisfying
\beqa
&&\big[{\textsf W}_0,{\textsf W}_{k+1}\big]=\big[{\textsf W}_{-k},{\textsf W}_{1}\big]=\frac{1}{(q^{1/2}+q^{-1/2})}\big({\tilde{\textsf G}_{k+1} } - {{\textsf G}_{k+1}}\big)\ ,\nonumber\\
&&\big[{\textsf W}_0,{\textsf G}_{k+1}\big]_q=\big[{\tilde{\textsf G}}_{k+1},{\textsf W}_{0}\big]_q=\rho{\textsf W}_{-k-1}-\rho{\textsf W}_{k+1}\ ,\nonumber\\
&&\big[{\textsf G}_{k+1},{\textsf W}_{1}\big]_q=\big[{\textsf W}_{1},{\tilde{\textsf G}}_{k+1}\big]_q=\rho{\textsf W}_{k+2}-\rho{\textsf W}_{-k}\ ,\nonumber\\
&&\big[{\textsf W}_0,{\textsf W}_{-k}\big]=0\ ,\quad 
\big[{\textsf W}_1,{\textsf W}_{k+1}\big]=0\ ,\quad \label{qOns}
\eeqa
and
\beqa
&&\big[{\textsf G}_{k+1},{\textsf G}_{l+1}\big]=0\ ,\quad   \big[{\tilde{\textsf G}}_{k+1},\tilde{{\textsf G}}_{l+1}\big]=0\ ,\quad
\big[{\tilde{\textsf G}}_{k+1},{\textsf G}_{l+1}\big]
+\big[{{\textsf G}}_{k+1},\tilde{{\textsf G}}_{l+1}\big]=0\ ,\nonumber
\eeqa
with fixed scalar $\rho$ and $k,l\in {\mathbb N}$. Finite dimensional representations are obtained, and examples of quantum integrable systems (XXZ spin chain, Sine-Gordon and Liouville field theories) which enjoy this symmetry are given. More generally, our framework opens the possibility of analyzing a large class of quantum integrable models from a new point of view, in the spirit of Onsager's approach \cite{Ons44}.
The paper is organized as follows. In Section 2, the fundamental relations (\ref{qOns}) are derived using its relation with a class of quadratic algebra, namely the reflection equation. Indeed, finite dimensional representations of its fundamental generators are shown to be generated from general solutions of the reflection equation. In particular, the closure of the algebra is ensured by the existence of a set of linear relations among the generators. Explicit expressions of mutually commuting quantities in terms of ${\textsf W}_{-k}$, ${\textsf W}_{k+1}$, ${\textsf G}_{k+1}$, ${\tilde{\textsf G}}_{k+1}$, generalizing (\ref{comOns}) are obtained. Also, we argue that similarly to the undeformed case, the integrable structure generated from our ``$q-$deformed'' Onsager's algebra coincides with the deformed Dolan-Grady integrable structure recently discovered in \cite{qDG, TriDiag}. In particular, we exhibit the correspondence in the simplest cases. In the last Section, we give some examples of quantum integrable systems which enjoy this (in)finite dimensional symmetry.

\section{Structure of the algebra}
In the last thirty years, one of the most important progress in the approach of quantum integrable systems
has been based on the star-triangle relations which originated in \cite{Ons44,Wannier} and led to the Yang-Baxter equations,
the theory of quantum groups as well as the quantum inverse scattering method. Although the star-triangle relations
 and the Onsager's algebra (\ref{Ons}) first appeared in the work of Onsager, to our knowledge, a direct and explicit
link between both structure has never been found or even noticed. Based on the recent results in \cite{qDG,TriDiag},
 in this Section we will exhibit such a link, relating a class of quadratic algebras - the reflection equation sometimes 
called the boundary Yang-Baxter equation - and a new finite dimensional algebra with deformation parameter $q$ generalizing
the Onsager's one.\footnote{As we will see later on, defining relations for the Onsager's algebra (\ref{Ons}) are 
recovered setting $q=1$.}. This link will allow us to derive the defining relations (\ref{qOns}) for the new algebra, as well as mutually commuting operators written in the basis of its fundamental generators.

\subsection{Fundamental generators and recursion relations}
Following \cite{qDG,TriDiag}, let us consider the quadratic algebra (reflection equation) which was first
introduced by Cherednik in \cite{Cher84}:
\beqa R(u/v)\ (K(u)\otimes I\!\!I)\ R(uv)\ (I\!\!I \otimes K(v))\
= \ (I\!\!I \otimes K(v))\ R(uv)\ (K(u)\otimes I\!\!I)\ R(u/v)\ .
\label{RE} \eeqa
This equation arises, for instance, in the context of quantum
integrable systems with boundaries \cite{Skly88}. We report the
reader to the literature on the subject for more details. For our
purpose, we restrict our attention to the trigonometric $R-$matrix
$R(u)$ which solves the Yang-Baxter equation. In the
spin$-\frac{1}{2}$ representation of
$U_{q^{1/2}}(\widehat{sl_2})$, it reads
\beqa R(u)=\sum_{i,j\in \{0,3,\pm\}}\omega_{ij}(u)\
\sigma_i\otimes\sigma_j\ , \label{R} \eeqa
where
\beqa \omega_{00}(u)&=&\frac{1}{2}(q^{1/2}+1)(u-q^{-1/2}u^{-1})\
,\qquad
\omega_{33}(u)= \frac{1}{2}(q^{1/2}-1)(u+q^{-1/2}u^{-1})\ ,\nonumber \\
\omega_{+-}(u)&=&\omega_{-+}(u)=q^{1/2}-q^{-1/2}\ ,\nonumber\eeqa
and $\sigma_j$ are Pauli matrices, $\sigma_\pm=(\sigma_1\pm
i\sigma_2)/2$. Suppose that one knows an ``initial'' two-dimensional matrix solution $K^{(0)}(u)$ of (\ref{RE}), then 
a family of solutions to (\ref{RE}) can be easily obtained using the so-called ``dressing'' procedure \cite{Skly88}.
Indeed, consider the fundamental solution (called $L-$operator) $L(u)$ of the quantum Yang-Baxter algebra
\beqa R(u/v)\big({L}(u)\otimes
{L}(v)\big)=\big({L}(v)\otimes{L}(u)\big)R(u/v)\ \label{RLL}\ .
 \eeqa
In the basis $\{S_\pm,s_3\}$ of the quantum algebra $U_{q^{1/2}}(sl_2)$ with defining relations
$[s_3,S_\pm]=\pm S_\pm$ and 
$[S_+,S_-]=(q^{s_3}-q^{-s_3})/(q^{1/2}-q^{-1/2})$\ , 
the $L-$operator takes the simple form:
\beqa {L}(u) =\left(
\begin{array}{cc}
 uq^{{1\over 4}}q^{s_3/2}- u^{-1}q^{-{1\over 4}}q^{-s_3/2}    &(q^{1/2}-q^{-1/2})S_-\\
(q^{1/2}-q^{-1/2})S_+    &  uq^{{1\over 4}}q^{-s_3/2}- u^{-1}q^{-{1\over 4}}q^{s_3/2}  \\
\end{array} \right) \ .\label{Lax} \eeqa
From the results of \cite{Skly88}, it follows,
for any parameter $v\in \mathbb{C}$, that
\beqa K^{(N)}(u)\equiv L_{\verb"N"}(uv)\cdot\cdot\cdot
L_{\verb"1"}(uv)K^{(0)}(u)L_{\verb"1"}(uv^{-1})\cdot\cdot\cdot
L_{\verb"N"}(uv^{-1})\ \label{Kdressed}\eeqa
acting on the quantum space ${ V}^{(N)}\equiv
\bigotimes_{{\verb"j"}={\verb"1"}}^{\verb"N"} { V}_{\verb"j"}$ also solves (\ref{RE}). Due to the choice (\ref{Lax}), it is clear that 
this general ``dressed'' solution  is a two-dimensional matrix in the auxiliary space with operator entries. Then, in full generality we decide to write it as
\beqa K^{(N)}(u) = \sum_{j\in\{0,3,\pm\}} \ \sigma_j \otimes
\Omega^{(N)}_j(u)\ ,\label{KN}\eeqa
where $\Omega^{(N)}_{j}(u)$ are rather complicated operators acting on the quantum space $V^{(N)}$. Our main objective
is now to write these operators in a more convenient form. In the following, we choose the trivial solution of (\ref{RE}) 
to be $K^{(0)}(u)\equiv (\sigma_+/c_0 + \sigma_-)/(q^{1/2}-q^{-1/2})$. Note that, according to (\ref{Lax}), (\ref{Kdressed}), the operators $\Omega_j^{(N)}(u)$ are combinations of Laurent polynomials of degree $-2N\leq d \leq 2N$ in the spectral parameter $u$ and operators acting solely on $\bigotimes_{{\verb"j"}={\verb"1"}}^{\verb"N"} U_{q^{1/2}}(sl_2)$.

\subsubsection{Case $N=1$}
For simplicity, let us start by considering $N=1$ in (\ref{Kdressed}). Note that this special case was first studied in details in \cite{Zab96}, having some interesting applications in the context of quasi-exactly solvable systems (Azbel-Hofstadter one in particular). This simplest ``dressed'' solution takes the form (\ref{KN}) for $N=1$ where the operators $\Omega^{(1)}_{j}(u)$ are easily written as \cite{qDG,TriDiag}:
\beqa \Omega^{(1)}_0(u)+\Omega^{(1)}_3(u)&=& uq^{1/2}{\textsf
W}^{(1)}_0 -u^{-1}q^{-1/2}{\textsf
W}^{(1)}_1 \ , \nonumber\\
\Omega^{(1)}_0(u)-\Omega^{(1)}_3(u)&=& uq^{1/2}{\textsf W}^{(1)}_1
-u^{-1}q^{-1/2}{\textsf W}^{(1)}_0\ ,\nonumber \\
\Omega^{(1)}_+(u)&=& \quad   \frac{q^{1/2} u^2 + q^{-1/2}
u^{-2}}{c_0(q^{1/2}-q^{-1/2})}
 +\frac{{\textsf G}^{(1)}_{1}}{q^{1/2}+q^{-1/2}} + \omega^{(1)}_0\ ,\nonumber \\
\Omega^{(1)}_-(u)&=&\frac{q^{1/2} u^2 + q^{-1/2}
u^{-2}}{{(q^{1/2}-q^{-1/2})}} +\frac{c_0 {\tilde{\textsf G}}^{(1)}_{1}}{q^{1/2}+q^{-1/2}}
+c_0\omega^{(1)}_0 \ ,\label{solfin1}
\eeqa
where the generators 
${\textsf W}^{(1)}_0,{\textsf W}^{(1)}_1,{\textsf G}^{(1)}_{1},{\tilde{\textsf G}}^{(1)}_{1}$ 
have been introduced. As shown in \cite{qDG}, the generators ${\textsf G}^{(1)}_{1},{\tilde{\textsf G}}^{(1)}_{1}$ are given by
${\textsf G}^{(1)}_{1}=[{\textsf W}^{(1)}_1,{{\textsf W}^{(1)}_0}]_q$ and ${\tilde{\textsf G}}^{(1)}_{1}=[{\textsf W}^{(1)}_0,{{\textsf W}^{(1)}_1}]_q$ where the $q-$commutator
\beqa
[X,Y]_{q}=q^{1/2}XY-q^{-1/2}YX\nonumber
\eeqa
has been introduced. In the basis of $U_{q^{1/2}}(sl_2)$, they admit the following representations:
\beqa
{\textsf W}^{(1)}_0&=& \frac{1}{c_0}vq^{1/4}S_+q^{s_3/2} + v^{-1}q^{-1/4}S_-q^{s_3/2}\ ,\nonumber\\ 
{\textsf W}^{(1)}_1&=& \frac{1}{c_0}v^{-1}q^{-1/4}S_+q^{-s_3/2} + vq^{1/4}S_-q^{-s_3/2}\ ,\nonumber\\
{\textsf G}^{(1)}_1&=&
\frac{(q^{1/2}+q^{-1/2})}{c_0(q^{1/2}-q^{-1/2})}\Big[\frac{(v^2+v^{-2})}{(q^{1/2}+q^{-1/2})}w_0^{(j)}-(v^{2}q^{s_3}+v^{-2}q^{-s_3})\Big]
+(q-q^{-1})S_-^2\ ,\nonumber\\  
{\tilde{\textsf G}}^{(1)}_{1}&=&\frac{(q^{1/2}+q^{-1/2})}{c_0(q^{1/2}-q^{-1/2})}\Big[\frac{(v^2+v^{-2})}{(q^{1/2}+q^{-1/2})}w_0^{(j)}-(v^{2}q^{-s_3}+v^{-2}q^{s_3})\Big]
+\frac{1}{c_0^2}(q-q^{-1})S_+^2\ .\label{rep1}
\eeqa
The remaining constant term $\omega^{(1)}_0$ in (\ref{solfin1}) can be either obtained directly from (\ref{Kdressed}), or follows by plugging (\ref{KN}) with  the structure (\ref{solfin1}) and (\ref{rep1}) in (\ref{RE}). This last requirement imposes strong constraints on the generators, leading to the so-called Askey-Wilson relations \cite{Zhed92}. We report the reader to \cite{qDG} for details. In any case, one finds
\beqa
\omega^{(1)}_0=-\frac{(v^2+v^{-2})}{c_0(q-q^{-1})}w_0^{(j)}\label{omega1}\ ,
\eeqa
where the Casimir operator eigenvalue of $U_{q^{1/2}}(sl_2)$ reads $w_0^{(j)}=q^{j+1/2}+q^{-j-1/2}$.

\subsubsection{Case $N=2$}
For $N=2$, the calculations are more involved but the procedure being straightforward, we do not need to report the detailed analysis. Part of the results below can be found in \cite{TriDiag}, but written in a slightly different form. In this case, the solution of the reflection equation (\ref{RE}) is given by (\ref{KN}) for $N=2$ with (see also \cite{TriDiag})
\beqa
\Omega^{(2)}_0(u)+\Omega^{(2)}_3(u)&=& uq^{1/2}\big(P^{(2)}_{0}(u){\textsf
W}^{(2)}_0 + P^{(2)}_{-1}(u){\textsf
W}^{(2)}_{-1}\big) 
-u^{-1}q^{-1/2}\big(P^{(2)}_{0}(u){\textsf
W}^{(2)}_1 + P^{(2)}_{-1}(u){\textsf
W}^{(2)}_{2}\big)\ ,\nonumber\\ 
\Omega^{(2)}_0(u)-\Omega^{(2)}_3(u)&=& uq^{1/2}
\big(P^{(2)}_{0}(u){\textsf
W}^{(2)}_1 + P^{(2)}_{-1}(u){\textsf
W}^{(2)}_{2}\big)
-u^{-1}q^{-1/2}
\big(P^{(2)}_{0}(u){\textsf
W}^{(2)}_0 + P^{(2)}_{-1}(u){\textsf
W}^{(2)}_{-1}\big) 
\ ,\nonumber\\ 
\Omega^{(2)}_+(u)&=& \quad   \frac{(q^{1/2} u^2 + q^{-1/2}
u^{-2})}{c_0(q^{1/2}-q^{-1/2})}P^{(2)}_{0}(u)
 +\frac{1}{q^{1/2}+q^{-1/2}}\big(P^{(2)}_{0}(u){\textsf G}^{(2)}_{1} +  P^{(2)}_{-1}(u){\textsf G}^{(2)}_{2}\big) + \omega^{(2)}_0\ ,\nonumber \\
\Omega^{(2)}_-(u)&=&\frac{(q^{1/2} u^2 + q^{-1/2}
u^{-2})}{{(q^{1/2}-q^{-1/2})}}P^{(2)}_{0}(u) +\frac{c_0}{q^{1/2}+q^{-1/2}}\big(P^{(2)}_{0}(u){\tilde{\textsf G}}^{(2)}_{1} +  P^{(2)}_{-1}(u){\tilde{\textsf G}}^{(2)}_{2}\big) 
+c_0\omega^{(2)}_0\ ,\label{solfin2}
\eeqa
where the generators ${\textsf W}^{(2)}_{-k},{\textsf W}^{(2)}_{k+1},{\textsf G}^{(2)}_{k+1},{\tilde{\textsf G}}^{(2)}_{k+1}$ for $k\in\{0,1\}$ have been introduced for further convenience. Their finite dimensional representations in the tensor product basis of $U_{q^{1/2}}(sl_2) \otimes U_{q^{1/2}}(sl_2)$ can be obtained as before, and coincide exactly with the ones corresponding to the general case $N$. So, we refer the reader to the next section for explicit expressions.  
Let us however mention that the exact expressions for the Laurent polynomials $P^{(2)}_{-k}(u)$ with $k\in\{0,1\}$ are given by
\beqa
P^{(2)}_{0}(u)&=& q^{1/2}u^2+q^{-1/2}u^{-2}+c_0(q^{1/2}-q^{-1/2})\omega_0^{(1)}-\frac{(v^{2}+v^{-2})}{(q^{1/2}+q^{-1/2})}w_0^{(j)}\ ,\nonumber\\
P^{(2)}_{-1}(u)&=& q^{1/2}+q^{-1/2}\ .\label{P2}
\eeqa
Finally, the constant term in (\ref{solfin2}) reads
\beqa
\omega^{(2)}_0=-\frac{(v^2+v^{-2})}{(q^{1/2}+q^{-1/2})}w_0^{(j)}\omega^{(1)}_0\ .\label{omega2}
\eeqa

\subsubsection{General case $N$}
For general values of $N$, we want to find an explicit expression for $K^{(N)}(u)$ such that the dependance on
the spectral parameter $u$ in 
the operators $\Omega^{(N)}_{j}(u)$ is disentangled, similarly to (\ref{solfin2}). 
Based on previous results for $N=1$ and $N=2$, we propose the following ansatz in (\ref{KN}) for $K^{(N)}(u)$:
\beqa
\Omega^{(N)}_0(u)+\Omega^{(N)}_3(u)&=& uq^{1/2}\sum_{k=0}^{N-1} P_{-k}^{(N)}(u)
{\textsf W}_{-k}^{(N)}-u^{-1}q^{-1/2}\sum_{k=0}^{N-1} P_{-k}^{(N)}(u) {\textsf W}_{k+1}^{(N)}\ ,\nonumber \\
\Omega^{(N)}_0(u)-\Omega^{(N)}_3(u)&=&uq^{1/2}\sum_{k=0}^{N-1} P_{-k}^{(N)}(u)
{\textsf W}_{k+1}^{(N)}-u^{-1}q^{-1/2}\sum_{k=0}^{N-1} P_{-k}^{(N)}(u) {\textsf W}_{-k}^{(N)}\ ,\nonumber\\
\Omega^{(N)}_+(u)&=&\frac{q^{1/2}u^2+q^{-1/2}u^{-2}}{c_0(q^{1/2}-q^{-1/2})}P_0^{(N)}(u)
+\frac{1}{q^{1/2}+q^{-1/2}}\sum_{k=0}^{N-1}P_{-k}^{(N)}(u){\textsf G}_{k+1}^{(N)}+{\omega}_0^{(N)}\ ,\nonumber\\
\Omega^{(N)}_-(u)&=&\frac{q^{1/2}u^2+q^{-1/2}u^{-2}}{(q^{1/2}-q^{-1/2})}P_0^{(N)}(u)
+\frac{c_0}{q^{1/2}+q^{-1/2}}\sum_{k=0}^{N-1}P_{-k}^{(N)}(u){\tilde
{\textsf G}}_{k+1}^{(N)}+c_0{\omega}_0^{(N)}\ ,\label{KansatzN}
\eeqa
where $P_{-k}^{(N)}(u)$ are Laurent polynomials to be determined. As explained above, according to the analysis of 
\cite{Skly88}, one knows that 
\beqa
K^{(N+1)}(u)\equiv L_{{\verb"N"}+{\verb"1"}}(uv)K^{(N)}(u)L_{ {\verb"N"}+{\verb"1"} }(uv^{-1})\label{decomp}
\eeqa
is also a solution of (\ref{RE}) provided $L(u)$ obeys (\ref{RLL}). For the ansatz (\ref{KansatzN}) to be correct for 
all $N$, we thus need to show that $K^{(N+1)}(u)$ keeps the form (\ref{KansatzN}) with $N\rightarrow N+1$. 
To do that, we proceed as follows. First, let us assume (\ref{KansatzN}) given $N$ fixed.
Then, we have to find an explicit relation between the tensor product of the ``old'' basis 
${\textsf W}^{(N)}_{-k},{\textsf W}^{(N)}_{k+1},{\textsf G}^{(N)}_{k+1},{\tilde{\textsf G}}^{(N)}_{k+1}$ 
with $U_{q^{1/2}}(sl_2)$ and the ``new'' one 
${\textsf W}^{(N+1)}_{-k},{\textsf W}^{(N+1)}_{k+1},{\textsf G}^{(N+1)}_{k+1},{\tilde{\textsf G}}^{(N+1)}_{k+1}$.
Such relation can be obtained, however being rather complicated we report the reader to Appendix A for the explicit recursion relations (\ref{repN+1N}) in the $(N+1)-$tensor product basis of $U_{q^{1/2}}(sl_2)$. At the same time, the following recursion relations for the Laurent polynomials:
\beqa
P_{0}^{(N+1)}(u)&=& \Big(q^{1/2}u^2+q^{-1/2}u^{-2} -\frac{(v^{2}+v^{-2})}{(q^{1/2}+q^{-1/2})}w_0^{(j)}\Big)P_{0}^{(N)}(u)
+c_0(q^{1/2}-q^{-1/2})\omega_0^{(N)}\ ,\nonumber\\
P_{-k}^{(N+1)}(u)&=& (q^{1/2}+q^{-1/2})P_{-k+1}^{(N)}(u)-P_{-k}^{(N)}(u)\frac{(v^{2}+v^{-2})}{(q^{1/2}+q^{-1/2})}w_0^{(j)}
\ \quad \mbox{for}\qquad k\in\{1,...,N-1\}\ ,\nonumber\\
P_{-N}^{(N+1)}(u)&=& (q^{1/2}+q^{-1/2})P_{-N+1}^{(N)}(u)\ \label{Precursion}
\eeqa
appear naturally. In addition, the constant term transforms as
\beqa
\omega_0^{(N+1)}=-\frac{(v^2+v^{-2})}{(q^{1/2}+q^{-1/2})}w_0^{(j)}\omega_0^{(N)} \ .\label{omegaN+1}
\eeqa
Remarkably, in the basis (\ref{repN+1N}) the operators $\Omega_j^{(N+1)}(u)$ 
can be drastically simplified. Indeed, using the ansatz (\ref{KansatzN}) and  the recursion relations for the polynomials
$P^{(N)}_{-k}(u)$ as written above, these operators reduce to
\beqa
\Omega^{(N+1)}_0(u)+\Omega^{(N+1)}_3(u)&=&\big(\Omega^{(N)}_0(u)+\Omega^{(N)}_3(u)\big)|_{N\rightarrow N+1} 
+ uq^{1/2}q^{s_3}\otimes \Delta_{1}^{(N+1)}- u^{-1}q^{-1/2}q^{-s_3}\otimes \Delta_{2}^{(N+1)}\ ,\nonumber \\
\Omega^{(N+1)}_0(u)-\Omega^{(N+1)}_3(u)&=&\big(\Omega^{(N)}_0(u)-\Omega^{(N)}_3(u)\big)|_{N\rightarrow N+1} 
+ uq^{1/2} q^{-s_3}\otimes \Delta_{2}^{(N+1)} - u^{-1}q^{-1/2}q^{s_3}\otimes \Delta_{1}^{(N+1)} \ ,\nonumber \\
\Omega^{(N+1)}_+(u)&=&\Omega^{(N)}_+(u)|_{N\rightarrow N+1}+(q^{1/2}-q^{-1/2})\Gamma^{(N+1)}_+\ ,\nonumber\\
\Omega^{(N+1)}_-(u)&=&\Omega^{(N)}_-(u)|_{N\rightarrow N+1}+(q^{1/2}-q^{-1/2})\Gamma^{(N+1)}_-\ ,\label{N+1parN}
\eeqa
where we denote
\beqa
\Gamma^{(N+1)}_+&=& vq^{-1/4}S_-q^{s_3/2}\otimes \Delta_{1}^{(N+1)} + v^{-1}q^{1/4}S_-q^{-s_3/2}\otimes \Delta_{2}^{(N+1)}
+\frac{1}{(q-q^{-1})}I\!\!I\otimes \Delta_{3}^{(N+1)}\ ,\nonumber\\
\Gamma^{(N+1)}_-&=& v^{-1}q^{1/4}S_+q^{s_3/2}\otimes \Delta_{1}^{(N+1)} + vq^{-1/4}S_+q^{-s_3/2}\otimes \Delta_{2}^{(N+1)}
+\frac{c_0}{(q-q^{-1})}I\!\!I\otimes \Delta_{4}^{(N+1)}\ .\nonumber
\eeqa
It is clear that the extra (unwanted) operators $\Delta_{j}^{(N+1)}$ with $j\in\{1,...,4\}$ (written below) must be 
vanishing in order for the ansatz (\ref{KansatzN}) to apply in the case $N+1$. To show that, let us focus 
for instance on $\Delta_{1}^{(N+1)}$. According to previous analysis, one has
\beqa
\Delta_{1}^{(N+1)}&=& -c_0(q^{1/2}-q^{-1/2})\omega_0^{(N)}{\textsf W}^{(N)}_{0}-P_{-N+1}^{(N)}(u){\textsf W}^{(N)}_{-N}\nonumber\\
&&\ \ \ +\sum_{k=1}^{N-1}
\Big((q^{1/2}u^2+q^{-1/2}u^{-2})P_{-k}^{(N)}(u)-(q^{1/2}+q^{-1/2})P_{-k+1}^{(N)}(u)\Big){\textsf W}^{(N)}_{-k}\ .
\label{deltaN+1}
\eeqa
In other words, the fundamental generators 
${\textsf W}^{(N)}_{-k},{\textsf W}^{(N)}_{k+1},{\textsf G}^{(N)}_{k+1},{\tilde{\textsf G}}^{(N)}_{k+1}$ must satisfy non-trivial linear relations. Furthermore, for consistency reasons the relations (\ref{deltaN+1}) must be independent of the spectral parameter $u$. It is then useful to notice that the Laurent polynomials (\ref{Precursion}) can be written as
\beqa
P_{-k+1}^{(N)}(u)=-\frac{1}{(q^{1/2}+q^{-1/2})}
\sum_{n=k-1}^{N-1}\Big(\frac{q^{1/2}u^2+q^{-1/2}u^{-2}}{q^{1/2}+q^{-1/2}}\Big)^{n-k+1}C_{-n}^{(N)} \label{PN}
\eeqa
for $1\leq k\leq N$, together with the initial 
condition $P_0^{(1)}(u)\equiv 1$ . Here the coefficient $C_{-k}^{(N)}$ are found to satisfy the recursion relations
\beqa
C_{0}^{(N)}&=& -(q-q^{-1})c_0\omega_0^{(N-1)} - \frac{(v^2+v^{-2})}{(q^{1/2}+q^{-1/2})}w_0^{(j)}C_0^{(N-1)}\ ,\nonumber\\
C_{-k}^{(N)}&=& (q^{1/2}+q^{-1/2})C_{-k+1}^{(N-1)}- \frac{(v^2+v^{-2})}{(q^{1/2}+q^{-1/2})}w_0^{(j)}C_{-k}^{(N-1)}
\ \quad\mbox{for}\quad 1\leq k\leq N-2\ ,\nonumber\\
C_{-N+1}^{(N)}&=& (q^{1/2}+q^{-1/2})C_{-N+2}^{(N-1)}\ \label{coeffC}
\eeqa
with $C^{(1)}_0=-(q^{1/2}+q^{-1/2})$. In particular, using (\ref{PN}) we immediately deduce for $k\in\{1,...,N-1\}$
\beqa
(q^{1/2}u^2+q^{-1/2}u^{-2})P_{-k}^{(N)}(u)-(q^{1/2}+q^{-1/2})P_{-k+1}^{(N)}(u) = C_{-k+1}^{(N)}\ .\label{simp}
\eeqa
Note that relations analogous to (\ref{deltaN+1}) also hold for 
$\Delta_{j}^{(N+1)}$ with $j\in\{2,3,4\}$ substituting 
${\textsf W}^{(N)}_{-k}$ by ${\textsf W}^{(N)}_{k+1}$, ${\textsf G}^{(N)}_{k+1}$ or ${\tilde{\textsf G}}^{(N)}_{k+1}$,
respectively. Then, as explained above,  $K^{(N+1)}(u)$ will have the structure (\ref{KansatzN}) provided 
$\Delta_{j}^{(N+1)}=0$ for $j\in\{1,...,4\}$
are satisfied for all values of $N$. After replacing  (\ref{simp}) in (\ref{deltaN+1}), we finally get the following 
(spectral parameter independent) linear relations among the fundamental generators
\beqa
&&c_0(q^{1/2}-q^{-1/2})\omega_0^{(N)}{\textsf W}^{(N)}_{0}
-\sum_{k=1}^{N} C_{-k+1}^{(N)}{\textsf W}^{(N)}_{-k}=0\ ,\nonumber\\
&&c_0(q^{1/2}-q^{-1/2})\omega_0^{(N)}{\textsf W}^{(N)}_{1}
-\sum_{k=1}^{N} C_{-k+1}^{(N)}{\textsf W}^{(N)}_{k+1}=0\ ,\nonumber\\
&&c_0(q^{1/2}-q^{-1/2})\omega_0^{(N)}{\textsf G}^{(N)}_{1}
-\sum_{k=1}^{N} C_{-k+1}^{(N)}{\textsf G}^{(N)}_{k+1}=0\ ,\nonumber\\
&&c_0(q^{1/2}-q^{-1/2})\omega_0^{(N)}{\tilde{\textsf G}}^{(N)}_{1}
-\sum_{k=1}^{N} C_{-k+1}^{(N)}{\tilde{\textsf G}}^{(N)}_{k+1}=0\ 
\label{linearrel}
\eeqa
with (\ref{omegaN+1}), (\ref{omega1}) and
\beqa
C_{-k+1}^{(N)}= (q^{1/2}+q^{-1/2})^{k}(-1)^{N-k+1}\Big(\frac{(v^2+v^{-2})}{(q^{1/2}+q^{-1/2})}w_0^{(j)}\Big)^{N-k}\frac{N!}{(k)!(N-k)!}\ \nonumber
\eeqa
for $k\in\{1,...,N\}$.
Using the 
representation (\ref{repN+1N}), we have checked explicitly that these linear relations are satisfied for all $N$. 
For simplicity, details are reported in Appendix B. The terms $\Delta^{(N+1)}_{j}$ vanishing in (\ref{N+1parN}),
it follows that $K^{(N+1)}(u)$ is given by (\ref{KN}), with
(\ref{KansatzN}) using the substitution $N\rightarrow N+1$. This being true for $N=1$ and $N=2$ as shown in previous sections,
we conclude that the general solutions of the reflection equations can be written as (\ref{KN}) with (\ref{KansatzN}) 
for all values of $N$, where the algebraic structure is now  encoded in the fundamental generators
${\textsf W}^{(N)}_{-k},{\textsf W}^{(N)}_{k+1},{\textsf G}^{(N)}_{k+1},{\tilde{\textsf G}}^{(N)}_{k+1}$.

\subsection{Integrable structure and generating function}
Applied to quantum integrable systems on the lattice, the generalized 
quantum inverse scattering approach provides a powerful method in order to derive in a 
systematic way a family of independent mutually commuting quantities.
For instance, as shown in \cite{Skly88} one can introduce the functional
\beqa
t^{(N)}(u) = tr_{0}\{K_+(u)K^{(N)}(u)\}
\label{tN}
\eeqa
with (\ref{KN}) and $K_{+}(u)$ which solves the ``dual'' reflection equation\,\footnote{The ``dual'' reflection equation 
follows from (\ref{RE}) by changing $u\rightarrow
q^{-1/2}u^{-1}$, $v\rightarrow q^{-1/2}v^{-1}$ and $K(u)$ in its transpose.}. Here $tr_{0}$ denotes the trace over the two-dimensional
auxiliary space. Then, it is proven in \cite{Skly88} that (\ref{tN}) satisfies
\beqa
\big[t^{(N)}(u),t^{(N)}(v)\big]=0\ \qquad \mbox{for all} \quad u,v\in {\mathbb C}\ ,\label{commut}
\eeqa
i.e. $t^{(N)}(u)$ constitutes the generating fonction for the mutually
commuting operators. For instance, let us plug the $c-$number solution of the ``dual'' reflection equation \cite{DV94,GZ94}
\beqa
{K}_+(u) =\left(
\begin{array}{cc}
 uq^{{1/2}}\kappa + u^{-1}q^{-{1/2}}\kappa^*    & \kappa_-(q^{1/2}+q^{-1/2})(qu^2-q^{-1}u^{-2})/c_0\\
\kappa_+(q^{1/2}+q^{-1/2})(qu^2-q^{-1}u^{-2}) &   uq^{{1/2}}\kappa^* + u^{-1}q^{-{1/2}}\kappa  \\
\end{array} \right) \ ,\label{K+}
\eeqa
with $\kappa^*\equiv \kappa^{-1}$, $\kappa_\pm$ arbitrary complex parameters and (\ref{KN}) with (\ref{KansatzN}) in 
(\ref{tN}). Immediately, we derive
\beqa t^{(N)}(u) =\sum_{k=0}^{N-1}(qu^2-q^{-1}u^{-2})P_{-k}^{(N)}(u)\ {\cal I}_{2k+1}^{(N)}\ + 
\ {\cal F}(u)\ I\!\!I\ , \label{expandt}\eeqa
with (\ref{PN}) and
\beqa
{\cal F}(u)=\frac{(q^{1/2}+q^{-1/2})(qu^2-q^{-1}u^{-2})}{c_0}\Big(\frac{(q^{1/2}u^2+q^{-1/2}u^{-2})}{(q^{1/2}-q^{-1/2})}P_0^{(N)}(u) + c_0\omega_0^{(N)}\Big)
\big(\kappa_+ +\kappa_-\big)\ .\nonumber
\eeqa
Note that the function ${\cal F}(u)$ is obviously not important from
an algebraic point of view. Here, we have introduced the operators  ${\cal I}_{2k+1}^{(N)}$ which can be written in terms of the fundamental generators as
\beqa
{\cal I}_{2k+1}^{(N)}=\kappa {\textsf W}^{(N)}_{-k} + \kappa^* {\textsf W}^{(N)}_{k+1} + \kappa_+ {\textsf G}^{(N)}_{k+1} 
+ \kappa_- {\tilde{\textsf G}}^{(N)}_{k+1}\ ,\label{Icons} 
\eeqa
for $k\in\{0,1,...,N-1\}$.  In particular,  the property (\ref{commut}) leads to 
\beqa
\big[{\cal I}_{2k+1}^{(N)},{\cal I}_{2l+1}^{(N)}\big]=0 \qquad \mbox{for all}\quad k,l\in\{0,...,N-1\}\ .\label{comI}
\eeqa
These latter commutation relations impose strong constraints on the fundamental generators.
Indeed, plugging (\ref{Icons}) in (\ref{comI}) one gets the commutation relations
\beqa
\big[{\textsf W}^{(N)}_{-k},{\textsf W}^{(N)}_{-l}\big]&=&0\ ,\quad   
\big[{\textsf W}^{(N)}_{k+1},{\textsf W}^{(N)}_{l+1}\big]=0\ ,\quad     
[{\textsf W}^{(N)}_{-k},{\textsf W}^{(N)}_{l+1}\big]+[{\textsf W}^{(N)}_{k+1},{\textsf W}^{(N)}_{-l}\big]=0\ ,\nonumber\\
\big[{\textsf G}^{(N)}_{k+1},{\textsf G}^{(N)}_{l+1}\big]&=&0\ ,\quad   
\big[{\tilde{\textsf G}}^{(N)}_{k+1},\tilde{{\textsf G}}^{(N)}_{l+1}\big]=0\ ,\quad 
\ \big[{\tilde{\textsf G}}^{(N)}_{k+1},{\textsf G}^{(N)}_{l+1}\big]
+\big[{{\textsf G}}^{(N)}_{k+1},\tilde{{\textsf G}}^{(N)}_{l+1}\big]=0\ ,\nonumber\\
\big[{\textsf W}^{(N)}_{k+1},{\textsf G}^{(N)}_{l+1}\big]&+&\big[{\textsf G}^{(N)}_{k+1},{\textsf W}^{(N)}_{l+1}\big]=0\ ,\qquad \quad
\big[{\textsf W}^{(N)}_{k+1},{\tilde{\textsf G}}^{(N)}_{l+1}\big]+\big[{\tilde{\textsf G}}^{(N)}_{k+1},{\textsf W}^{(N)}_{l+1}\big]=0\ ,\nonumber\\
\big[{\textsf W}^{(N)}_{-k},{\textsf G}^{(N)}_{l+1}\big]&+&\big[{\textsf G}^{(N)}_{k+1},{\textsf W}^{(N)}_{-l}\big]=0\ ,\qquad \quad
\big[{\textsf W}^{(N)}_{-k},{\tilde{\textsf G}}^{(N)}_{l+1}\big]+\big[{\tilde{\textsf G}}^{(N)}_{k+1},{\textsf W}^{(N)}_{-l}\big]=0\ .\label{opcom}
\eeqa
A few remarks can now be done.  First, although from the beginning we have considered 
tensor product representations of $U_{q^{1/2}}(sl_2)$, the form of (\ref{Icons}) esentially relies on
the structure (\ref{KN}) with (\ref{KansatzN}). Then, 
classifying all possible finite, infinite dimensional or cyclic tensor product representations different from
(\ref{repN+1N}) is an interesting open question.
Secondly, it should be stressed that given $N$, the relations (\ref{linearrel}) and their generalizations (see Appendix B (\ref{c1})-(\ref{c4})) are responsible of the truncation of the integrable hierarchy, i.e. any quantity ${\cal I}_{2k+1}^{(N)}$ for $k\geq N$ can be written in terms of 
all ${\cal I}_{2k+1}^{(N)}$ with $k\leq N-1$. It follows that given $N$, there are only $N-$independent mutually commuting fundamental 
quantities. To conclude, let us mention that for the special case $N=1$ (and $k=0$) it is easy to check that (\ref{Icons}) 
coincides exactly with the result of \cite{Zab96}.

\subsection{Fundamental $q-$deformed commutation relations}
We are now interested in the algebraic structure associated with the fundamental generators ${\textsf W}^{(N)}_{-k},{\textsf W}^{(N)}_{k+1},{\textsf G}^{(N)}_{k+1},{\tilde{\textsf G}}^{(N)}_{k+1}$. 
The form of (\ref{KansatzN}) and the representations (\ref{repN+1N}) being determined by the quadratic algebra (\ref{RE}), this equation fixes all fundamental relations among the generators. To extract these relations, we proceed as follows. First, from (\ref{Lax}) and (\ref{R}), one has
\beqa
\pi^{(1/2)}\big[L(u)\big]=R(u)\ ,
\eeqa 
where $\pi^{(1/2)}$ denotes the spin$-\frac{1}{2}$ representation of $U_{q^{1/2}}(sl_2)$. Then, plugging $K^{(N)}(u)$ in (\ref{RE}), this latter equation can be written as
\beqa
&&(\pi^{(1/2)}\times id^{(N)})\big[L_{{\verb"N"}+{\verb"1"}}(uv^{-1})K^{(N)}(u)L_{ {\verb"N"}+{\verb"1"} }(uv)\big]\big(I\!\!I \otimes K^{(N)}(v)\big)\nonumber\\
&&\qquad \qquad \qquad \qquad \qquad =\big(I\!\!I \otimes K^{(N)}(v)\big)(\pi^{(1/2)}\times id^{(N)})\big[L_{{\verb"N"}+{\verb"1"}}(uv)K^{(N)}(u)L_{ {\verb"N"}+{\verb"1"} }(uv^{-1})\big]\ .\label{inter}
\eeqa
Being satisfied for any value of the spectral parameter $u$, following \cite{qDG} we can consider its asymptotic expansion for $u\rightarrow \infty$. Replacing (\ref{KN}) with (\ref{KansatzN}) in (\ref{inter}), one finds that
the leading equation is trivially satisfied. However, the (next) two subleading ones read
\beqa
(\pi^{(1/2)}\times id^{(N)})\big[{\textsf W}_0^{(N+1)}\big]|_{v\rightarrow v^{-1}}K^{(N)}(v) &=&  (\pi^{(1/2)}\times id^{(N)})\big[{\textsf W}_0^{(N+1)}\big]K^{(N)}(v)\ ,\nonumber\\
(\pi^{(1/2)}\times id^{(N)})\big[{\textsf W}_1^{(N+1)}\big]|_{v\rightarrow v^{-1}}K^{(N)}(v) &=&  (\pi^{(1/2)}\times id^{(N)})\big[{\textsf W}_1^{(N+1)}\big]K^{(N)}(v)\ .\label{interfin}
\eeqa
Using the recursion relations (\ref{repN+1N}) for the finite dimensional representations of the fundamental generators and $\pi^{(1/2)}[S_\pm]=\sigma_\pm$ and $\pi^{(1/2)}[s_3]=\sigma_3/2$, one has
\beqa
\pi^{(1/2)}\big[{\textsf W}_0^{(N+1)}\big] =\left(
\begin{array}{cc}
 q^{{1\over 2}}{\textsf W}_0^{(N)}    & v/c_0\\
v^{-1}   &  q^{-{1\over 2}}{\textsf W}_0^{(N)}  \\
\end{array} \right) \ ,\qquad \pi^{(1/2)}\big[{\textsf W}_1^{(N+1)}\big] =\left(
\begin{array}{cc}
 q^{-{1\over 2}}{\textsf W}_1^{(N)}   & v^{-1}/c_0\\
v   &  q^{{1\over 2}}{\textsf W}_1^{(N)}  \\
\end{array} \right) \ .\nonumber\eeqa
It is now easy to simplify the intertwinning relations (\ref{interfin}). After some calculations, the constraints (\ref{linearrel}) appear naturally with the use of (\ref{simp}). These latter relations being satisfied (see Appendix B for details), ommiting the index $N$ we end up with the defining relations (\ref{qOns}) for all $k\in\{0,...,N-1\}$ provided one identifies
\beqa \rho=\frac{(q^{1/2}+q^{-1/2})^2}{c_0}\ .\label{valrho}\eeqa
Note that some of the relations (\ref{qOns}) already appear in (\ref{opcom}). 
Actually, it is easy to give an alternative derivation of the $q-$deformed commutation relations (\ref{qOns}) using the representation (\ref{repN+1N}). Indeed, for any $N$ the constraints (\ref{opcom}) are satisfied. Let us consider $N\rightarrow N+1$ in these constraints, and  replace the generators ${\textsf W}^{(N+1)}_{-k}$, ${\textsf W}^{(N+1)}_{k+1}$, ${\textsf G}^{(N+1)}_{k+1}$, ${\tilde{\textsf G}}^{(N+1)}_{k+1}$, $k\in\{0,...,N\}$ by 
their finite dimensional representations (\ref{repN+1N}). After some straightforward calculations, the 
$q-$deformed relations (\ref{qOns}) arise explicitely. Consequently, this shows perfect consistency between the approach
(\ref{inter}) associated with the symmetries underlying (\ref{RE}), and the properties of the integrable structure (\ref{Icons}), i.e. (\ref{opcom}). It follows that all elements ${\textsf W}^{(N)}_{-k},{\textsf W}^{(N)}_{k+1},{\textsf G}^{(N)}_{k+1},{\tilde{\textsf G}}^{(N)}_{k+1}$ for $k\in\{0,...,N-1\}$ are generated from ${\textsf W}^{(N)}_{0},{\textsf W}^{(N)}_{1}$  using the recursion relations  (\ref{qOns}).

\subsection{Relation with tridiagonal algebras and deformed Dolan-Grady hierarchy}
The integrable structure (\ref{Ons}) is known to be identical \cite{Perk} with the Dolan-Grady construction introduced and studied in \cite{DG}, but corresponding to a different notation. This latter structure was found to apply to the class of Hamiltonian of the form
\beqa
{\cal H} = \kappa A_0 + \kappa^* A_1 \ .\label{HDG}
\eeqa
As shown in \cite{DG}, the integrability condition of the related models (Ising, XY,...) relies on the existence of two (necessary and sufficient) conditions, the ``Dolan-Grady relations'', defined by
\beqa
[A_0,[A_0,[A_0,A_1]]]=16[A_0,A_1]\qquad \mbox{and}\qquad [A_1,[A_1,[A_1,A_0]]]=16[A_1,A_0]\ .\label{relDG}
\eeqa
All higher mutually commuting quantities beyond (\ref{HDG}) can be written solely in terms of the fundamental operators $A_0,A_1$. Note that provided $A_0,A_1$ satisfy (\ref{relDG}), the whole Onsager's algebra (\ref{Ons}) is generated. Also,  $A_0,A_1$ as well as the other generators of the Onsager's algebra can be expressed in the basis of the loop algebra $\tilde{sl_2}$ \cite{Davies,Roan91,Date00}. We refer the reader to these works for details.\vspace{1mm}

Surprisingly, a ``$q-$deformed'' analogue of the Dolan-Grady relations (\ref{relDG}) recently appeared in the context  of  $P-$ and $Q-$polynomial association schemes \cite{Ter93,Ter01,Ter03}:
\beqa \big[{\textsf A},\big[{\textsf A},\big[{\textsf A},{\textsf
A}^*\big]_{q}\big]_{q^{-1}}\big]=  \rho\big[{\textsf A},{\textsf A}^*\big]\ \qquad \mbox{and}\qquad
\big[{\textsf A}^*,\big[{\textsf A}^*,\big[{\textsf A}^*,{\textsf
A}\big]_q\big]_{q^{-1}}\big]=  \rho\big[{\textsf A}^*,{\textsf A}\big]\ .
 \label{relqDG}\eeqa
By [\cite{Ter03}, Definition 3.9] the {\it tridiagonal algebra} $\mathbb{T}$ is the associative algebra with unity generated by two symbols ${\textsf A},{\textsf A}^*$ subject to the relations (\ref{relqDG}). We call ${\textsf A},{\textsf A}^*$ the {\it standard generators}. Here $q$ is a deformation parameter (usually assumed to be not a root of unity) and $\rho$ is a fixed scalar. Let ${\cal V}$ denote a finite dimensional irreducible module for $\mathbb{T}$. Then the pair of linear transformations ${\textsf
A}:{\cal V}\rightarrow {\cal V}$ and ${\textsf A}^*:{\cal V}\rightarrow {\cal V}$ is said to be a
tridiagonal (TD) pair [\cite{Ter01}, Definition 1.1], which complete classification remains an open problem.
The subset of TD pairs such that
${\textsf A},{\textsf A}^*$ have eigenspaces of dimension one is called Leonard pairs,
classified in \cite{TerLP01}. In particular, Leonard pairs satisfy (for details, see \cite{TerAW03}) the
so-called Askey-Wilson (AW) relations first introduced by Zhedanov
in \cite{Zhed92}. Other examples of TD pairs can be found in
\cite{Ter03,TerIto04}: for $\rho=0$ in which case (\ref{relqDG})
reduce to $q-$Serre relations; for $q=1$ and $\rho=16$ which leads
to the Dolan-Grady relations (\ref{relDG}). The more general situation $\rho\neq0$, $q\neq1$
was recently considered in details in \cite{qDG,TriDiag}. There, it was found that TD pairs ${\textsf A},{\textsf A}^*$
admit a realization in terms of the quantum affine Kac-Moody algebra
$U_{q^{1/2}}(\widehat{sl_2})$. This algebra is generated by $Q_\pm,{\overline Q}_\pm$ and $H$
subjects to
\beqa &&q^{-1/2}Q_\pm{\overline Q}_\pm - q^{1/2}{\overline Q}_\pm
Q_\pm =0\ ,\nonumber\\
&&q^{1/2}Q_\pm{\overline Q}_\mp - q^{-1/2}{\overline Q}_\mp
Q_\pm =\frac{q^{\pm 2H}-1}{q^{1/2}-q^{-1/2}}\ ,\nonumber\\
&&q^{\epsilon H}Q_\pm = q^{\pm \epsilon}Q_\pm q^{\epsilon H}\
,\quad q^{\epsilon H}{\overline Q}_\pm = q^{\pm
\epsilon}{\overline Q}_\pm q^{\epsilon H}\ \label{Uqsl2}\eeqa
together with the $q-$Serre relations
\beqa Q_\pm^3 Q_\mp -(1+q+q^{-1})Q_\pm^2Q_\mp Q_\pm +
(1+q+q^{-1})Q_\pm Q_\mp Q_\pm^2 - Q_\mp Q_\pm^3 =0\ ,\nonumber \\
{\overline Q}_\pm^3 {\overline Q}_\mp -(1+q+q^{-1}){\overline
Q}_\pm^2{\overline Q}_\mp {\overline Q}_\pm +
(1+q+q^{-1}){\overline Q}_\pm {\overline Q}_\mp {\overline
Q}_\pm^2 - {\overline Q}_\mp {\overline Q}_\pm^3 =0\ .\nonumber
\eeqa
Also, the Hopf algebraic structure of $U_{q^{1/2}}(\widehat{sl_2})$ is ensured by the
coproduct $\Delta: U_{q^{1/2}}(\widehat{sl_2}) \rightarrow
U_{q^{1/2}}(\widehat{sl_2}) \times U_{q^{1/2}}(\widehat{sl_2})$
associated with (\ref{Uqsl2}) acting on the fundamental generators as
\beqa \Delta(Q_\pm)&=& Q_\pm \otimes I\!\!I + q^{\pm H} \otimes
Q_\pm \
,\nonumber\\
 \Delta({\overline Q}_\pm)&=& {\overline Q}_\pm \otimes I\!\!I +
q^{\mp H} \otimes {\overline Q}_\pm \
,\nonumber\\
\Delta(q^{H})&=& q^H \otimes q^{H}\ .\label{coproduct}\eeqa
More generally, one defines the $N-$coproduct $\Delta^{(N)}: \
U_{q^{1/2}}(\widehat{sl_2}) \longrightarrow
U_{q^{1/2}}(\widehat{sl_2}) \otimes \cdot\cdot\cdot \otimes
U_{q^{1/2}}(\widehat{sl_2})$ as
\beqa \Delta^{(N)}\equiv (id\times \cdot\cdot\cdot \times id
\times \Delta)\circ \Delta^{(N-1)}\ \nonumber\eeqa
for $N\geq 3$ with $\Delta^{(2)}\equiv \Delta$,
$\Delta^{(1)}\equiv id$. Note that the opposite $N-$coproduct
$\Delta'^{(N)}$ is similarly defined with $\Delta'\equiv \sigma
\circ\Delta$,  where the permutation map $\sigma(x\otimes y
)=y\otimes x$ for all $x,y\in U_{q^{1/2}}(\widehat{sl_2})$ is
used. As noticed in \cite{TriDiag}, it is not difficult to show that some linear combinations 
of $U_{q^{1/2}}(\widehat{sl_2})$ generators satisfy (\ref{relqDG}). More generally, using the homorphism property of 
(\ref{coproduct}), it is straightforward to check that
\beqa
{\textsf A}\equiv\Delta^{(N)}\Big(\frac{1}{c_0}Q_+ + {\overline Q}_- + \epsilon_+ q^{H}\Big) \qquad
\mbox{and} \qquad {\textsf A}^*\equiv\Delta^{(N)}\Big(Q_- + \frac{1}{c_0}{\overline
Q}_+ + \epsilon_-q^{-H}\Big)\ \label{TDinit}
\eeqa
defines a family of TD pairs (i.e. (\ref{relqDG}) is satisfied) for arbitrary parameters $\epsilon_\pm$ and the identification (\ref{valrho}).
Note that although the special case $\epsilon_\pm=0$ is considered in most of this paper, as pointed out in \cite{TriDiag} the algebraic structure remains the same for $\epsilon_\pm\neq 0$. The only difference arises in the exact expressions for the $c-$number coefficients in the linear relations (\ref{linearrel}).\vspace{1mm}  

A $q-$deformed analogue of the Dolan-Grady integrable structure \cite{DG} was proposed in \cite{qDG,TriDiag}, also based on the properties of the quadratic algebra (\ref{RE}). In the basis of ${\textsf A},{\textsf A}^*$, 
the two first charges simply read
\beqa {\cal I}_1 &=& \kappa{\textsf A} +
\kappa^*{\textsf A}^* \ ,\nonumber\\
{\cal I}_3 &=& \kappa\Big(\frac{c_0}{(q^{1/2}+q^{-1/2})^2}\big[\big[{\textsf
A},{{\textsf A}^*}\big]_q,{\textsf A}\big]_{q} +
{{\textsf A}^*}\Big)+ \kappa^*\Big(\frac{c_0}{(q^{1/2}+q^{-1/2})^2}\big[\big[{{\textsf
A}^*},{{\textsf A}}\big]_q,{{\textsf A}^*}\big]_{q}
 + {{\textsf A}}\Big) \ , \label{IN}\eeqa
which are mutually commuting in virtue of (\ref{relqDG}). Comparison between the results of previous Sections and the ones of \cite{TriDiag} can be done easily, and allow us to find the explicit expression of the fundamental generators ${\textsf W}^{(N)}_{-k},{\textsf W}^{(N)}_{k+1},{\textsf G}^{(N)}_{k+1},{\tilde{\textsf G}}^{(N)}_{k+1}$ in terms of ${\textsf A},{\textsf A}^*$ satisfying (\ref{relqDG}) in the simplest cases $N=1$ and $N=2$. Furthermore, it provides an alternative check of the fundamental $q-$deformed relations (\ref{qOns}), the linear relations (\ref{linearrel}) as well as their generalizations (\ref{c1})-(\ref{c4}). 

\subsubsection{Case $N=1$}
Assuming that the entries $\Omega_j^{(1)}(u)$ are combinations of Laurent polynomials of degree $-2\leq d\leq 2$ in $u$ and operators, it was shown in \cite{qDG} that any solution $K^{(1)}(u)$ takes the form (\ref{KN}) for $N=1$ and the identification
\beqa
&&{\textsf W}^{(1)}_{0}= {\textsf A}\ ,\quad {\textsf W}^{(1)}_{1}= {\textsf A}^*\ ,\quad 
{\textsf G}^{(1)}_{1}= \big[{\textsf A}^*,{\textsf A}\big]_q\ ,\quad {\tilde{\textsf G}}^{(1)}_{1}
=\big[{\textsf A},{\textsf A}^*\big]_{q}\ .\label{id1}
\eeqa
It should be stressed that this does {\it not} require the relation (\ref{decomp}) to be satisfied, and only relies on
the degree of the Laurent polynomials in $\Omega_j^{(1)}(u)$. Nevertheless,
the fact that $K^{(1)}(u)$ satisfies (\ref{RE}) imposes strong constraints on the generators ${\textsf A},{\textsf A}^*$:
plugging (\ref{KN}) for $N=1$ with (\ref{solfin1}) and using (\ref{id1}), one obtains the so-called Askey-Wilson relations \cite{Zhed92}
\beqa {{\textsf A}}^2  {{\textsf A}^*} +{{\textsf
A}^*}{{\textsf A}}^2-(q+q^{-1}){{\textsf
A}}{{\textsf A}^*}{{\textsf A}}&=&\frac{(q^{1/2}+q^{-1/2})^2}{c_0}{{\textsf A}^*}-\frac{(v^2+v^{-2})}{c_0}w_0^{(j)}{{\textsf A}}\ ,\nonumber\\
{{\textsf A}^*}^2  {\textsf A} +{\textsf
A} {{\textsf A}^*}^2-(q+q^{-1}){\textsf A}^*{\textsf A} {\textsf
A}^*&=& \frac{(q^{1/2}+q^{-1/2})^2}{c_0}{\textsf A} - \frac{(v^2+v^{-2})}{c_0}w_0^{(j)} {\textsf
A}^*\  .\label{AW} \eeqa
In particular, it is easy to check these relations for the representation (\ref{rep1}). Having in mind the $q-$deformed
relations (\ref{qOns}) and the corresponding hierarchy (\ref{Icons}) for $\kappa_\pm=0$, it is natural to propose from (\ref{IN}) the following identification:
\beqa
{\textsf W}^{(1)}_{-1}&\equiv&\frac{c_0}{(q^{1/2}+q^{-1/2})^2}\big[{{\textsf A},\big[{\textsf
A}^*,{\textsf A}\big]_q\big]}_{q} +
{{\textsf A}^*}\ ,\nonumber\\
{\textsf W}^{(1)}_{2}&\equiv&\frac{c_0}{(q^{1/2}+q^{-1/2})^2}\big[\big[{{\textsf
A}^*},{{\textsf A}}\big]_q,{{\textsf A}^*}\big]_{q}
 + {{\textsf A}}\ .\label{idW1}
\eeqa
This notation introduced in (\ref{AW}), together with (\ref{id1}) leads to a set of very simple linear relations among the fundamental generators. These relations are actually responsible of the truncation of the hierarchy (\ref{IN}) for $N=1$, i.e. one finds that ${\cal I}_3$ is proportional to ${\cal I}_1$. More generally, any higher charge ${\cal I}_{2k+1}$ for $k\geq 1$ can be expressed in terms of the first one \cite{TriDiag}. Based on this truncation which occurs at $N=1$ and the defining relations (\ref{qOns}), one finds a slightly generalized version of the first two equations in (\ref{linearrel}) for the special case $N=1$:
\beqa
c_0(q^{1/2}-q^{-1/2})\omega_0^{(1)} {\textsf W}^{(1)}_{-l}-C_{0}^{(1)}{\textsf W}^{(1)}_{-l-1}=0\ \qquad \mbox{and}\qquad
c_0(q^{1/2}-q^{-1/2})\omega_0^{(1)} {\textsf W}^{(1)}_{l+1}-C_{0}^{(1)}{\textsf W}^{(1)}_{l+2}=0\ \label{Wrel1} 
\eeqa
for any $l\geq 0$. In particular, the AW relations correspond to $l=0$.

\subsubsection{Case $N=2$}
Based on the results of \cite{TriDiag}, we proceed similarly. Assuming now that the entries $\Omega_j^{(1)}(u)$ are combinations of Laurent polynomials of degree $-4\leq d\leq 4$ in $u$ and operators, the solution $K^{(2)}(u)$ takes the form (\ref{KN}) for $N=2$ with the identification
\beqa
&&{\textsf W}^{(2)}_{0}= {\textsf A}\ ,\qquad \ \qquad {\textsf W}^{(2)}_{-1}=  \frac{c_0}{(q^{1/2}+q^{-1/2})^2}\big[{{\textsf A},\big[{\textsf
A}^*,{\textsf A}\big]_q\big]}_{q} + {{\textsf A}^*}\ ,\label{id2}\\
&&{\textsf W}^{(2)}_{1}= {\textsf A}^*\ ,\qquad \qquad W^{(2)}_{2}=
\frac{c_0}{(q^{1/2}+q^{-1/2})^2}\big[\big[{{\textsf
A}^*},{{\textsf A}}\big]_q,{{\textsf A}^*}\big]_{q}
 + {{\textsf A}}\ .\nonumber\\
&&{\textsf G}^{(2)}_{1}= \big[{\textsf A}^*,{\textsf A}\big]_q\ ,\qquad {\textsf G}^{(2)}_{2}= \alpha_1\big[{{\textsf A}^*}^2,{{\textsf A}}^2\big]_{q^2} + \alpha_2\big[{{\textsf A}^*},{{\textsf A}}\big]^2_{q} + \alpha_3\big[{{\textsf A}^*},{{\textsf A}}\big]^{2} + \frac{(q^{1/2}-q^{-1/2})}{(q+q^{-1})}({\textsf A}^2+{{\textsf A}^*}^2)+ \alpha_0\ ,\nonumber\\
&&{\tilde{\textsf G}}^{(2)}_{1}=\big[{\textsf A},{\textsf A}^*\big]_{q}\ 
,\qquad {\tilde{\textsf G}}^{(2)}_{2}= \alpha_1\big[{{\textsf A}}^2,{{\textsf A}^*}^2\big]_{q^2} + \alpha_2\big[{{\textsf A}},{{\textsf A}^*}\big]^2_{q} + \alpha_3\big[{{\textsf A}},{{\textsf A^*}}\big]^{2} + \frac{(q^{1/2}-q^{-1/2})}{(q+q^{-1})}({\textsf A}^2+{{\textsf A}^*}^2)+ \alpha_0\ ,\nonumber
\eeqa
where 
\beqa
\alpha_0&=&\frac{2}{c_0(q^{1/2}-q^{-1/2})}\left(\frac{w_0^{(j)\,2}}{(q^{1/2}+q^{-1/2})^2}-1\right)\left(1-\frac{(v^4+v^{-4})}{(q+q^{-1})}\right)\ ,\nonumber\\
\alpha_1&=&-\frac{c_0(q^{1/2}-q^{-1/2})}{(q^2-q^{-2})}\ ,\quad \alpha_2=\frac{c_0(q+q^{-1})}{(q-q^{-1})(q^{1/2}+q^{-1/2})}\ ,\quad \alpha_3=-\frac{c_0(q^{1/2}+q^{-1/2})}{(q^2-q^{-2})}\
.\nonumber
\eeqa
It is important to notice that the expressions of the fundamental generators ${\textsf W}^{(2)}_{0},{\textsf W}^{(2)}_{1},{\textsf G}^{(2)}_{1},{\tilde{\textsf G}}^{(2)}_{1}$  in terms of
${\textsf A},{\textsf A}^*$ remains unchanged compared to the case $N=1$. In addition, the proposal
(\ref{idW1}) is confirmed. Plugging $K^{(2)}(u)$ given by (\ref{KN}) with (\ref{solfin2}) in  (\ref{RE}) and
using (\ref{id2}), we obtain the $N=2$ generalization of the AW relations (\ref{AW}):
\beqa
\big[{{\textsf A},{{\textsf G}}^{(2)}_{2}]}_{q}&=&
\frac{2(v^2+v^{-2})}{(q^{1/2}+q^{-1/2})^2}w_0^{(j)}[{\textsf A},[{\textsf A}^\ast,{\textsf A}]_q]_q
+[{\textsf A}^\ast,[{\textsf A}^\ast,{\textsf A}]_q]_{q^{-1}} +\frac{2(v^2+v^{-2})}{c_0}w_0^{(j)}{\textsf A}^\ast\nonumber\\
&&\qquad -\left(\frac{(q^{1/2}+q^{-1/2})^2}{c_0}
+\frac{(v^2+v^{-2})^2}{c_0(q^{1/2}+q^{-1/2})^2}w_0^{(j)\,2}\right){\textsf A} \ ,
\nonumber\\
\big[{{\textsf G}}^{(2)}_{2},{\textsf A}^*]_{q}&=&\frac{2(v^2+v^{-2})}{(q^{1/2}+q^{-1/2})^2}w_0^{(j)}[{\textsf A}^\ast,[{\textsf A},{\textsf A}^\ast]_q]_q
+[{\textsf A},[{\textsf A},{\textsf A}^\ast]_q]_{q^{-1}}+\frac{2(v^2+v^{-2})}{c_0}w_0^{(j)}{\textsf A}\nonumber\\
&&\qquad -\left(\frac{(q^{1/2}+q^{-1/2})^2}{c_0}
+\frac{(v^2+v^{-2})^2}{c_0(q^{1/2}+q^{-1/2})^2}w_0^{(j)\,2}\right){\textsf A}^\ast \ .
\label{AW2}
\eeqa
Notice that $\big[{\tilde{\textsf G}}^{(2)}_{2},{{\textsf A}]}_{q}=\big[{{\textsf A},{{\textsf G}}^{(2)}_{2}]}_{q}$ and
$\big[{{\textsf G}}^{(2)}_{2},{{\textsf A}^*]}_{q}=\big[{{\textsf A}^*,{\tilde{\textsf G}}^{(2)}_{2}]}_{q}$, so that no other relations than (\ref{AW2}) are obtained. Based on the algebraic structure (\ref{qOns}), as before it is then natural to propose
\beqa
{\textsf W}^{(2)}_{-2}&\equiv&\frac{c_0}{(q^{1/2}+q^{-1/2})^2}\big[{{\textsf A},{{\textsf G}}^{(2)}_{2}}\big]_{q} + {\textsf W}^{(2)}_2
\ ,\nonumber\\
{\textsf W}^{(2)}_{3}&\equiv&\frac{c_0}{(q^{1/2}+q^{-1/2})^2}\big[{{\textsf G}}^{(2)}_{2},{{\textsf A}^*}\big]_{q}
 + {\textsf W}^{(2)}_{-1}\ ,\label{idW2}
\eeqa
where the  explicit expression of ${{\textsf W}}^{(2)}_{2},{{\textsf W}}^{(2)}_{-1},{{\textsf G}}^{(2)}_{2}$ in terms of ${{\textsf A}},{{\textsf A}^*}$ are used in the r.h.s of (\ref{idW2}). Identifying (\ref{idW2}) in the constraint (\ref{AW2}), one finds immediately the linear relations (\ref{linearrel}) for $N=2$. Using the same argument as before about the truncation of the hierarchy, it follows that only ${\cal I}_1$ and ${\cal I}_3$ are independent for $N=2$. All higher charges being linear combinations of them, we deduce
\beqa
c_0(q^{1/2}-q^{-1/2})\omega_0^{(2)} {\textsf W}^{(2)}_{-l}-C_{0}^{(2)}{\textsf W}^{(2)}_{-l-1}-C_{-1}^{(2)}{\textsf W}^{(2)}_{-l-2}&=&0\ ,\nonumber\\
c_0(q^{1/2}-q^{-1/2})\omega_0^{(2)} {\textsf W}^{(2)}_{l+1}-C_{0}^{(2)}{\textsf W}^{(2)}_{l+2}-C_{-1}^{(2)}{\textsf W}^{(2)}_{l+3}&=&0\ \label{Wrel2}
\eeqa
for any $l\geq 0$. For $l=0$, the relations (\ref{AW2}) are recovered.

\subsubsection{General case $N$}
For more general values of $N$, although technically difficult it is well expected that the fundamental generators can be written solely in terms of ${\textsf A},{\textsf A}^*$. Indeed, for $q=1$ the deformed Dolan-Grady integrable structure must reduce to the undeformed one, so that operators in the two hierarchies are in one-to-one correspondence.
This goes beyond the scope of this paper, so we do not pursue the analysis for $N>2$.\vspace{1mm}

We now want to focus our attention on the construction of linear relations generalizing (\ref{linearrel}), similarly to (\ref{Wrel1}) and (\ref{Wrel2}). First, it is an exercise to show using (\ref{Wrel1}) and (\ref{qOns}) that the relation
\beqa
c_0(q^{1/2}-q^{-1/2})\omega_0^{(1)} {\textsf G}^{(1)}_{l+1}-C_{0}^{(1)}{\textsf G}^{(1)}_{l+2}=0\ \label{Grel1} 
\eeqa
is satisfied, and similarly for ${\tilde{\textsf G}}^{(1)}_{l+1}$. Indeed, this is in agreement with the argument based on the truncation of the hierarchy (\ref{Icons}) for the more general case $\kappa_\pm\neq 0$.
Any charge ${\cal I}^{(N)}_{2k+1}$ for $k\geq N$ being expressed as a linear combination of all charges ${\cal I}^{(N)}_{2k+1}$ for $k\leq N-1$, and the parameters $\kappa,\kappa^*,\kappa_\pm$ being independent, we propose
for general values of $N$ the following relations generalizing (\ref{linearrel})
\beqa
&&c_0(q^{1/2}-q^{-1/2})\omega_0^{(N)}W_{-l}^{(N)}-\sum^{N}_{k=1}C_{-k+1}^{(N)}W_{-k-l}^{(N)}=0\ ,\label{c1}\\
&&c_0(q^{1/2}-q^{-1/2})\omega_0^{(N)}W_{l+1}^{(N)}-\sum^{N}_{k=1}C_{-k+1}^{(N)}W_{k+l+1}^{(N)}=0\ ,\label{c2}\\
&&c_0(q^{1/2}-q^{-1/2})\omega_0^{(N)}G_{l+1}^{(N)}-\sum^{N}_{k=1}C_{-k+1}^{(N)}G_{k+l+1}^{(N)}=0\ ,\label{c3}\\
&&c_0(q^{1/2}-q^{-1/2})\omega_0^{(N)}{\tilde G}_{l+1}^{(N)}-\sum^{N}_{k=1}C_{-k+1}^{(N)}{\tilde G}_{k+l+1}^{(N)}=0\ ,\label{c4}\
\eeqa
for any $l\geq 0$. For $N=1$ and $N=2$, these relations were obtained above. We have checked explicitly that these relations also hold for general values of $N$, using the explicit finite dimensional representations of the generators. We report the reader to Appendix B for details. As a consistency check, let us mention that the relations (\ref{c3}), (\ref{c4}) actually follow from  (\ref{c1}) and (\ref{c2}), using the $q-$deformed commutation relations (\ref{qOns}).

\section{Concluding remarks}
The Onsager's algebra is known to be generated from two elements $A_0,A_1$ satisfying the Dolan-Grady relations (\ref{relDG}). Defining $4G_1=[A_1,A_0]$, all higher elements $A_k,G_l$ in (\ref{Ons}) are generated from the recursion relations \cite{Davies,Roan91,Date00} 
\beqa
A_{k+1}-A_{k-1}= \frac{1}{2}\big[G_1,A_k\big]\ ,\qquad G_l =\frac{1}{4}\big[A_l,A_0\big]\ .\label{recOns}
\eeqa
The relations (\ref{relDG}) are actually {\it sufficient} to reconstruct the Onsager's algebra (\ref{Ons}).
Furthermore, for finite dimensional representations the spectral properties of (\ref{HDG}) (as well as arbitrary combinations of $A_k,G_l$) are known to be encoded in the closure of the algebra \cite{Davies,Roan91} which reads (for some coefficients $\alpha_k$)
\beqa
\sum_{k}\alpha_{k}A_{k-l}=0\ ,\qquad \sum_{k}\alpha_{k}G_{k-l}=0\ .\label{closOns}
\eeqa

In this paper, based on the link between the quadratic algebra (\ref{RE}) and the deformed Dolan-Grady integrable structure recently discovered in \cite{qDG,TriDiag}, we have found that the algebra (\ref{Ons}) introduced by Onsager in \cite{Ons44} admits a $q-$deformed infinite dimensional analogue (\ref{qOns}) with fundamental generators
${\textsf W}_{-k},{\textsf W}_{k+1},{\textsf G}_{k+1},{\tilde{\textsf G}}_{k+1}$ with $k\in {\mathbb N}$. Similarly to the Onsager's algebra, the integrable structure follows from the ``$q-$deformed'' Dolan-Grady relations (i.e. tridiagonal algebra) (\ref{relqDG}) with ${\textsf A}\rightarrow {\textsf W}_0$, ${\textsf A}^*\rightarrow {\textsf W}_1$  and the ``$q-$deformed'' recursion relations in (\ref{qOns}).  
It should be stressed that this new algebra possesses either finite or infinite dimensional representations: Finite dimensional representations (\ref{repN+1N}) have been obtained, in which case the generators satisfy a set of linear relations generalizing the Askey-Wilson ones (\ref{c1})-(\ref{c4}). On the other hand, in the limit\,\footnote{Of course, the linear relations (\ref{c1})-(\ref{c4}) need to be clarified in this case.} $N\rightarrow \infty$ vertex operators representations can be used \cite{qDG}.\vspace{0.2cm} 

This new symmetry ensures the existence of an (in)finite number (associated with the (in)finite parameter $N$) of mutually commuting quantities given by (\ref{Icons}). For the special (undeformed) case $q=1$, the defining relations (\ref{relqDG}), (\ref{qOns}) coincide exactly with the ones considered in \cite{DG}, \cite{Davies}. Indeed, simple comparison between (\ref{comOns}), (\ref{qOns}) and (\ref{Icons}) gives the exact relation between our generators and the ones in \cite{Ons44}:
\beqa
&&{\textsf W}_{-k}|_{q=1} \equiv (A_{k}+ A_{-k})/2 \ , \qquad {\textsf W}_{k+1}|_{q=1} \equiv (A_{k+1}+ A_{-k+1})/2\ ,\nonumber\\
&&{\textsf G}_{k+1}|_{q=1}= -{\tilde{\textsf G}_{k+1}|_{q=1}}\equiv 4G_{k+1}\ .
\eeqa
Also, the linear relations (\ref{linearrel}) as well as their generalizations (\ref{c1})-(\ref{c4}) reduce to the ones proposed in \cite{Davies}.\vspace{2mm} 

The most interesting problem now is to analyze quantum integrable models with this new mathematical framework, in order to extract any nonperturbative information. In this direction, identifying the models with such underlying symmetry is obviously the first thing  to be done. In particular, it should be stressed that (\ref{qOns}) is closely related with $U_{q^{1/2}}(sl_2)$ with generators $Q_\pm$, ${\overline Q}_\pm$ \cite{qDG,TriDiag}. Then, it is well expected that the characteristics of the model are encoded in the parameters $N$, $q$, $c_0$, $v$ whereas the fundamental generators should correspond to some observables.  Below, we give various examples of quantum integrable models (lattice, massive, boundary or conformal) which enjoy the symmetry (\ref{qOns}).\vspace{0.2cm}

$\bullet$ {\bf XXZ open spin chain with general boundary conditions:} The fundamental generators ${\textsf W}^{(N)}_{0},{\textsf W}^{(N)}_{1}$ are related\,\footnote{Note that the nonlocal charges derived in \cite{Doik04} correspond to certain integrable boundary conditions, not the most general ones.} with the nonlocal conserved charges obtained in \cite{Doik04} using the method proposed in \cite{Nepo98,DelMac03}, and $N$ corresponds to the number of sites. The deformation parameter $q$ characterizes the anisotropy $\Delta=(q^{1/2}+q^{-1/2})/2$ of the model, whereas $\kappa$, $\kappa_\pm$ ($c_0$) are non-diagonal left (right) boundary conditions, respectively. Also, $v=1$.
Note that for more general right boundary conditions associated with extra parameters $\epsilon_\pm$, the algebraic structure remains essentially unchanged. It follows that the underlying fundamental {\it symmetry} of the XXZ open spin chain with general boundary conditions {\it is} the $q-$deformed Onsager's algebra (\ref{qOns}) with representations (\ref{repN+1N}). This symmetry, sometimes called ``boundary quantum group algebra'', is in one-to-one correspondence with the tridiagonal algebra \cite{Ter93,Ter01,Ter03} as shown in \cite{TriDiag}. Based on previous analysis, it follows that the transfer matrix can be simply written as (\ref{expandt}). Details will be reported elsewhere.\vspace{0.2cm} 

$\bullet$ {\bf Sine-Gordon quantum field theory:} In the bulk, the sine-Gordon model is known to possess nonlocal conserved charges \cite{Ber91}, usually denoted $Q_\pm$, ${\overline Q}_\pm$, generating a $U_{q^{1/2}}(sl_2)$ symmetry. Actually, ${\textsf W}^{(N)}_{0},{\textsf W}^{(N)}_{1}$ for $N\rightarrow \infty$ admit a vertex operator representation in one-to-one correspondence with linear combinations of these charges \cite{qDG} and parametrized by $c_0$ (arbitrary). The deformation parameter $q$ and parameter $v$ are easily related with the coupling constant $\beta^2$ and the rapidity of the fundamental particles (soliton/antisoliton), respectively. In case of a non-dynamical \cite{GZ94} or a dynamical boundary \cite{BK} the model still remains integrable, but the symmetry is restricted. The corresponding nonlocal conserved charges have been constructed in \cite{Nepo98,DelMac03}, and generate an example of tridiagonal algebra \cite{qDG,TriDiag}. For these boundary integrable models, one has the identification  $N\rightarrow\infty$ and $c_0=1$.\vspace{0.2cm}
 
$\bullet$ {\bf Liouville quantum field theory:}  In this conformal limit of the sine-Gordon model, it is easy to check that either $Q_+,{\overline Q}_-$ or $Q_-,{\overline Q}_+$ are conserved (commuting with the stress-energy tensor). Then, an infinite number of conserved quantities are obtained from (\ref{Icons}) for $N\rightarrow \infty$ and some vanishing parameters $\kappa,\kappa^*,\kappa_\pm$. It follows that the Liouville field theory (as well as its boundary counterpart) enjoys the symmetry generated by a subalgebra of (\ref{qOns}).\vspace{0.2cm}

$\bullet$ {\bf Quasi-exactly solvable systems, Bethe ansatz and $q-$difference equations:} For general values of $N\neq1$, the spectral problem associated with (\ref{Icons}) leads to a system of
partial $q-$difference equations that clearly needs further investigation. For the special case $N=1$, one obtains a second-order $q-$difference equation which has been considered in details in \cite{Zab96,Wieg95}, and lead to Bethe equations. Interestingly, for the limit $q\rightarrow 1$ it becomes the Heun (or similarly the P${\ddot{o}}$schl-Teller) equation. Furthermore, at this special value of the deformation parameter the Onsager algebra (\ref{Ons}) exhibited and studied in the context of quasi-exactly solvable systems and nonlinear holomorphic supersymmetry \cite{Plyu} is recovered. Then, we expect our construction will provide a new (algebraic) approach to the surprising relation between conformal field theory and differential equations pointed out in \cite{BLZ}, as well as its massive counterpart.\vspace{0.2cm}

Related problems will be considered elsewhere.

\vspace{0.5cm}

\noindent{\bf Acknowledgements:} We thank P. Terwilliger for comments. K. Koizumi is supported by CNRS and French Ministry of Education and Research. Part of this work is supported by the TMR Network EUCLID ``Integrable models
and applications: from strings to condensed matter'', contract
number HPRN-CT-2002-00325.\vspace{1cm}

\centerline{\bf {\large Appendix A: Tensor product representations of the fundamental generators}}
\vspace{0.0cm}
\beqa
{\textsf W}_0^{(N+1)}&=& \frac{1}{c_0}vq^{1/4}S_+ q^{s_3/2}\otimes I\!\!I + v^{-1}q^{-1/4}S_- q^{s_3/2}\otimes
I\!\!I + q^{s_3}\otimes {\textsf W}_0^{(N)}\ ,\nonumber\\
{\textsf W}_1^{(N+1)}&=& \frac{1}{c_0}v^{-1}q^{-1/4}S_+ q^{-s_3/2}\otimes I\!\!I + vq^{1/4}S_- q^{-s_3/2}\otimes
I\!\!I + q^{-s_3}\otimes {\textsf W}_1^{(N)}\ ,\label{repN+1N}\\
{\textsf G}_{1}^{(N+1)}&=& (q-q^{-1})S_-^2\otimes
I\!\!I
-\frac{(q^{1/2}+q^{-1/2})}{c_0(q^{1/2}-q^{-1/2})} (v^{2}q^{s_3}+v^{-2}q^{-s_3}) \otimes I\!\!I
+I\!\!I \otimes {\textsf G}_{1}^{(N)}\nonumber\\
&&\ \ \ +(q-q^{-1})\left(
vq^{-1/4}S_-q^{s_3/2}\otimes {\textsf W}_0^{(N)}
+v^{-1}q^{1/4}S_-q^{-s_3/2}\otimes {\textsf W}_{1}^{(N)}
\right)
+\frac{(v^2+v^{-2})w_0^{(j)}}{c_0(q^{1/2}-q^{-1/2})}\otimes I\!\!I\ ,\nonumber\\
{\tilde{\textsf G}}_1^{(N+1)}&=&
\frac{(q-q^{-1})}{c_0^2}S_+^2\otimes I\!\!I
-\frac{(q^{1/2}+q^{-1/2})}{c_0(q^{1/2}-q^{-1/2})}(v^{2}q^{-s_3}+v^{-2}q^{s_3})\otimes I\!\!I
+ I\!\!I \otimes {\tilde {\textsf G}}_{1}^{(N)}\nonumber\\
&&\ \ \ +\frac{(q-q^{-1})}{c_0}\left(
v^{-1}q^{1/4}S_+q^{s_3/2}\otimes {\textsf W}_0^{(N)}
+vq^{-1/4}S_+q^{-s_3/2}\otimes {\textsf W}_{1}^{(N)}
\right)
+\frac{(v^{2}+v^{-2})w_0^{(j)}}{c_0(q^{1/2}-q^{-1/2})}\otimes I\!\!I\ ,\nonumber\\
\nonumber\\
{\textsf W}_{-k-1}^{(N+1)}&=&\frac{(w_0^{(j)}-(q^{1/2}+q^{-1/2})q^{s_3})}{(q^{1/2}+q^{-1/2})}\otimes
{\textsf W}_{k+1}^{(N)}
-\frac{(v^2+v^{-2})}{(q^{1/2}+q^{-1/2})}I\!\!I\otimes {\textsf W}_{-k}^{(N)}
+\frac{(v^2+v^{-2})w_0^{(j)}}{(q^{1/2}+q^{-1/2})^2}{\textsf W}_{-k}^{(N+1)}\nonumber\\
&&\ \ \ +\frac{(q^{1/2}-q^{-1/2})}{(q^{1/2}+q^{-1/2})^2}
\left(vq^{1/4}S_+q^{s_3/2}\otimes
{\textsf G}_{k+1}^{(N)}+c_0v^{-1}q^{-1/4}S_-q^{s_3/2}\otimes {\tilde {\textsf G}}_{k+1}^{(N)}\right)+q^{s_3}\otimes {\textsf W}_{-k-1}^{(N)}
\ ,\nonumber\\
\nonumber\\
{\textsf W}_{k+2}^{(N+1)}&=&\frac{(w_0^{(j)}-(q^{1/2}+q^{-1/2})q^{-s_3})}{(q^{1/2}+q^{-1/2})}\otimes
{\textsf W}_{-k}^{(N)}
-\frac{(v^2+v^{-2})}{(q^{1/2}+q^{-1/2})}I\!\!I\otimes {\textsf W}_{k+1}^{(N)}
+\frac{(v^2+v^{-2})w_0^{(j)}}{(q^{1/2}+q^{-1/2})^2}{\textsf W}_{k+1}^{(N+1)}
\nonumber\\
&&\ \ \ +\frac{(q^{1/2}-q^{-1/2})}{(q^{1/2}+q^{-1/2})^2}
\left(v^{-1}q^{-1/4}S_+q^{-s_3/2}\otimes
{\textsf G}_{k+1}^{(N)}+c_0vq^{1/4}S_-q^{-s_3/2}\otimes {\tilde {\textsf G}}_{k+1}^{(N)}\right)+q^{-s_3}\otimes {\textsf W}_{k+2}^{(N)}
\ ,\nonumber\\
\nonumber\\
{\textsf G}_{k+2}^{(N+1)}&=& 
\frac{c_0(q^{1/2}-q^{-1/2})^2}{(q^{1/2}+q^{-1/2})}
S_-^2\otimes {\tilde {\textsf G}}_{k+1}^{(N)}
-\frac{1}{(q^{1/2}+q^{-1/2})}(v^{2}q^{s_3}+v^{-2}q^{-s_3})\otimes {\textsf G}_{k+1}^{(N)} 
+I\!\!I \otimes {\textsf G}_{k+2}^{(N)}\nonumber\\
&&\ \ \ + (q-q^{-1})\left(
vq^{-1/4}S_-q^{s_3/2}\otimes \big({\textsf W}_{-k-1}^{(N)}-{\textsf W}_{k+1}^{(N)}\big)
+v^{-1}q^{1/4}S_-q^{-s_3/2}\otimes \big({\textsf W}_{k+2}^{(N)}-{\textsf W}_{-k}^{(N)}\big)
\right)\nonumber\\
&&\ \ \ +\frac{(v^2+v^{-2})w_0^{(j)}}{(q^{1/2}+q^{-1/2})^2}{\textsf G}_{k+1}^{(N+1)}\ ,\nonumber\\
\nonumber\\
{\tilde {\textsf G}}_{k+2}^{(N+1)}&=& 
\frac{(q^{1/2}-q^{-1/2})^2}{c_0(q^{1/2}+q^{-1/2})}
S_+^2\otimes {{\textsf G}}_{k+1}^{(N)}
-\frac{1}{(q^{1/2}+q^{-1/2})}(v^{2}q^{-s_3}+v^{-2}q^{s_3})\otimes {\tilde {\textsf G}}_{k+1}^{(N)} 
+I\!\!I \otimes {\tilde {\textsf G}}_{k+2}^{(N)}\nonumber\\
&&\ \ \ + \frac{(q-q^{-1})}{c_0}\left(
v^{-1}q^{1/4}S_+q^{s_3/2}\otimes \big({\textsf W}_{-k-1}^{(N)}-{\textsf W}_{k+1}^{(N)}\big)
+vq^{-1/4}S_+q^{-s_3/2}\otimes \big({\textsf W}_{k+2}^{(N)}-{\textsf W}_{-k}^{(N)}\big)
\right)\nonumber\\
&&\ \ \ +\frac{(v^2+v^{-2})w_0^{(j)}}{(q^{1/2}+q^{-1/2})^2}{\tilde{\textsf G}}_{k+1}^{(N+1)}\ \nonumber
\eeqa
for $k\in\{0,1,...,N-1\}$.
\vspace{1cm}

\centerline{\bf \large Appendix B: Generalized linear relations}
\vspace{3mm}
The purpose of this Appendix is to show that the linear relations (\ref{linearrel}), as well as their generalizations (\ref{c1})-(\ref{c4}), are satisfied for all values of $N$. For $N$ fixed,
let us first assume that $W^{(N)}_{-k},W^{(N)}_{k+1},G^{(N)}_{k+1}$ and ${\tilde G}^{(N)}_{k+1}$ with $k\in \{0,...,N\}$ satisfy (\ref{c1})-(\ref{c4}). Then, a straightforward calculation based on the finite dimensional representations (\ref{repN+1N}) shows that
\beqa
  &&\sum_{k=1}^{N+1} C^{(N+1)}_{-k+1}W_{-k-l}^{(N+1)}-c_0(q^{1/2}-q^{-1/2})
    \omega_{0}^{(N+1)}W_{-l}^{(N+1)}\nonumber\\
    &&\qquad = \frac{w_0^{(j)}-(q^{1/2}+q^{-1/2})q^{s_3}}{(q^{1/2}+q^{-1/2})}\otimes \sum_{k=1}^{N+1}\beta^{(N+1)}_{-k+1}W_{k+l}^{(N)}
    -\frac{v^2+v^{-2}}{q^{1/2}+q^{-1/2}}I\!\!I\otimes\sum_{k=1}^{N+1}\beta^{(N+1)}_{-k+1} W_{-k+1-l}^{(N)}
    \nonumber\\
    &&\qquad+\frac{(q^{1/2}-q^{-1/2})}{(q^{1/2}+q^{-1/2})}
        \left(vq^{1/4}S_+q^{s_3/2}\otimes \sum_{k=1}^{N+1}\beta^{(N+1)}_{-k+1}G_{k+l}^{(N)}+c_0v^{-1}q^{-1/4}S_-q^{s_3/2}\otimes\sum_{k=1}^{N+1}\beta^{(N+1)}_{-k+1} {\tilde G}_{k+l}^{(N)}\right)\nonumber\\
       &&\qquad+q^{s_3}\otimes \sum_{k=1}^{N+1}\beta^{(N+1)}_{-k+1}W_{-k-l}^{(N)}\nonumber\\ 
          \noalign{\vskip 0.1cm}
    &&\qquad+\frac{(v^2+v^{-2})w_0^{j}}{(q^{1/2}+q^{-1/2})^2}\beta_0^{(N+1)}W_{-l}^{(N+1)}-c_0(q^{1/2}-q^{-1/2})
    \omega_{0}^{(N+1)}W_{-l}^{(N+1)}\ ,\label{devN+1}
\eeqa
where $l\geq0$ and 
\beqa
\beta_{-k+1}^{(N+1)}=\sum_{m=k}^{N+1} \left(\frac{(v^2+v^{-2})w_0^{(j)}}{(q^{1/2}+q^{-1/2})^2}\right)^{m-k}
C^{(N+1)}_{-k+1}\ .\nonumber
\eeqa
Using (\ref{coeffC}), it is easy to notice that
\beqa
\beta_{-(k-1)}^{(N+1)}=(q^{1/2}+q^{-1/2})C_{-k+2}^{(N)}\quad {\rm\  for}\quad  2\leq k\leq N+1 \qquad \mbox{and}\qquad \beta_0^{(N+1)}=-c_0(q-q^{-1})\omega_0^{(N)}\ .\nonumber
\eeqa
Replacing these expressions and  (\ref{omegaN+1}) in (\ref{devN+1}), all terms vanish in virtue of (\ref{c1})-(\ref{c4}). Consequently,
\beqa
\sum_{k=1}^{N+1} C^{(N+1)}_{-k+1}W_{-k-l}^{(N+1)}-c_0(q^{1/2}-q^{-1/2})
\omega_{0}^{(N+1)}W_{-l}^{(N+1)}=0\ ,\label{cN+1}
\eeqa
provided (\ref{c1})-(\ref{c4}) for $N$ fixed. Similar relations also holds for
$W^{(N+1)}_{k+1},G^{(N+1)}_{k+1}$ and ${\tilde G}^{(N+1)}_{k+1}$ with $k\in \{0,...,N+1\}$. Now, we proceed by recursion:\\

$\bullet$ General case $N$ and $l=0$: the linear relations (\ref{linearrel}) are satisfied for $N=1$ and $N=2$, either corresponding to the Askey-Wilson relations (for $N=1$), or its $N=2$ generalization (\ref{AW2}). According to (\ref{cN+1}), it follows that (\ref{linearrel}) are satisfied for all values of $N$. \\

$\bullet$ General case $N$ and arbitrary $l\geq 0$:  Due to (\ref{linearrel}), (\ref{KN}) admits the representation (\ref{KansatzN}) and the mutually commuting operators take the form (\ref{Icons}). For $N=1$ and $N=2$, we obtained (\ref{Wrel1}) for arbitrary $l\geq 0$. Due to (\ref{cN+1}), (\ref{c1}) is indeed satisfied for all values of $N$.
Clearly, similar analysis can be done (see (\ref{Grel1})) for  $W^{(N)}_{k+1},G^{(N)}_{k+1}$ and ${\tilde G}^{(N)}_{k+1}$ with $k\in \{0,...,N\}$. Then, we conclude that (\ref{c1})-(\ref{c4}) are satisfied for all values of $N$.

\end{document}